\documentclass[12pt,preprint]{aastex}
\begin{document}

\title{Distributions of Gamma-Ray Bursts and Blazars in the $L_{\rm p}-E_{\rm p}$ Plane and Possible Implications for their Radiation Physics}
\author{Fen Lyu\altaffilmark{1,2}, En-Wei Liang\altaffilmark{1,3,4}, Yun-Feng Liang\altaffilmark{1}, Xue-Feng Wu\altaffilmark{2}, Jin Zhang\altaffilmark{4,3}, Xiao-Na Sun\altaffilmark{1}, Rui-Jing Lu\altaffilmark{1}, Bing Zhang\altaffilmark{5,1}
}
\altaffiltext{1}{Department of Physics and GXU-NAOC Center for Astrophysics
and Space Sciences, Guangxi University, Nanning 530004,
China; lew@gxu.edu.cn}
\altaffiltext{2}{Purple Mountain Observatory, Chinese Academy of Sciences, Nanjing, 210008, China; xfwu@pmo.ac.cn}
\altaffiltext{3}{Key Laboratory for the Structure and Evolution of Celestial Objects, CAS, Kuming 650011}
\altaffiltext{4}{National Astronomical Observatories, Chinese Academy of Sciences, Beijing, 100012, China}
\altaffiltext{5}{Department of Physics and Astronomy, University of
Nevada, Las Vegas, NV 89154; zhang@physics.unlv.edu}

\begin{abstract}
We present a spectral analysis for a sample of redshift known GRBs observed with {\em Fermi}/GBM. Together with the results derived from our systematical spectral energy distribution modeling with the leptonic models for a {\em Fermi}/LAT blazar sample, we compare the distributions of the GRBs and the blazars by plotting the synchrotron peak luminosity ($L_{\rm s}$) and the corresponding peak photon energy $E_{\rm s}$ of blazars in the $L_{\rm p}-E_{\rm p}$ plane of GRBs, where $L_{\rm p}$ and $E_{\rm p}$ are the peak luminosity and peak photon energy of the GRB time-integrated $\nu f_\nu$ spectrum, respectively. The GRBs are in the high-$L_{\rm p}$, high-$E_{\rm p}$ corner of the plane and a tight $L_{\rm p}-E_{\rm p}$ relation is found, i.e., $L_{\rm p}\propto E_{\rm p}^{2.13^{+0.54}_{-0.46}}$. Both FSRQs and LBLs are clustered in the low-$E_{\rm p}$, low-$L_{\rm p}$ corner. IBLs and HBLs have $E_{\rm s}\sim 2\times 10^{-3} - 10^{2}$ keV and $L_{\rm s} \sim 10^{44} - 10^{47}$ erg s$^{-1}$, but no dependence of $L_{\rm s}$ on $E_{\rm s}$ is found. We show that the tight $L_p-E_p$ relation of GRBs is potentially explained with the synchrotron radiation of fast-cooling electrons in a highly magnetized ejecta, and the weak anti-correlation of $L_{\rm s}-E_{\rm s}$ for FSRQs and LBLs may be attributed to synchrotron radiation of slow-cooling electrons in a moderately magnetized ejecta. The distributions of IBLs and HBLs in the $L_{\rm p}-E_{\rm p}$ plane may be interpreted with synchrotron radiation of fast-cooling electrons in a matter-dominated ejecta. These results may present a unified picture for the radiation physics of relativistic jets in GRBs and blazars within the framework of the leptonic synchrotron radiation models.
\end{abstract}


\section{Introduction}
It is believed that the radiation of both gamma-ray bursts (GRBs) and blazars are from relativistic jets powered by central black holes (BHs) in different mass scales. The central engines of GRBs may be newly born stellar BHs from either core collapses of massive stars (Woosley 1993; Paczy\'{n}ski 1998) or mergers of two compact objects (e.g. Eichler et al. 1989). The central engines of blazars, which are divided into BL Lac objects (BL Lacs) and flat spectrum radio quasars (FSRQs) according to the emission line features, are believed to be super-massive rotating BHs. It is speculated that the physics of jet launching and dissipation in different BH mass scales may be essentially the same (Mirabel 2004; Ghisellini 2005; Zhang 2007;  Ma et al. 2014). Some comparative studies show evidence of similar jet properties between blazars and GRBs (e.g., Wang \& Wei 2011; Wu et al. 2011; Nemmen et al. 2012; Zhang et al. 2013a; Wang et al. 2014).

Broad-band spectral energy distribution (SED) is critical to investigate radiation physics and jet properties. The broad-band SEDs of blazars usually show two humps, which can be well explained by synchrotron radiation and inverse Compton (IC) scattering of the relativistic electrons in the jets, respectively (e.g., Maraschi et al. 1992; Dermer et al. 1992; Ghisellini et al. 1996; Zhang et al. 2012a, 2013). The observed synchrotron peak energy ($E_{\rm s}$) of FSRQs is usually in the IR/optical band, whereas $E_{\rm s}$ of BL Lacs may range from IR to the X-ray band (frequency ranging from $10^{12}$ to $10^{18}$ Hz). The BL Lacs are classified into low-, intermediate-, and high- synchrotron peaking categories (LBLs, IBLs, and HBLs) based on the observed $E_{\rm s}$.

The SEDs of the prompt gamma-ray emission of most GRBs, typically from several keV to GeV as observed by the {\em Fermi} mission, are usually adequately fitted by an empirical smoothly broken power-law function (Abdo et al. 2009a; Zhang et al. 2011), the so-called ``Band function'' (Band et al. 1993). The peak energy $E_{\rm p}$ may vary from several keV to MeV. The spectra of a small fraction of GRBs, such as GRB 090510 (Ackermann et al. 2010) and GRB 090926 (Ackermann et al. 2011), show an extra power-law component or a bump in the {\em Fermi}/LAT band. The radiation physics of the Band function is a great mystery in the GRB field. Synchrotron emission is the leading model (e.g., M\'esz\'aros et al. 1994; Wang et al. 2009; Daigne et al. 2011; Zhang \& Yan 2011; Uhm \& Zhang 2014). Alternatively, the photosphere radiation (e.g., M{\'e}sz{\'a}ros et al. 2002; Rees \& M\'esz\'aros 2005; Pe'er et al. 2006; Giannios 2008; Ryde \& Pe'er 2009; Beloborodov 2010; Pe'er \& Ryde 2011; Vurm et al. 2013) and synchrotron self-Compton process (e.g., Racusin et al. 2008; Kumar \& Panaitescu 2008; cf. Piran et al. 2009; Resmi \& Zhang 2012) have been also suggested. The time resolved spectra of some GRBs, such as GRB 090902B (Abdo et al. 2009b; Ryde et al. 2010; Zhang et al. 2011) and GRB 081221 (Hou et al. 2014, in preparation) can be fitted with a quasi-thermal radiation model. Fan et al. (2012) suggested that some empirical relations in GRBs can be explained with the photosphere radiation model. On the other hand, the photosphere model also suffers some criticisms. For example, the low energy spectral index of photosphere emission is typically too hard (typically with a photon index $\sim$ +0.5) to interpret the observations (typically $\sim -1$), and it cannot interpret the commonly observed hard-to-soft evolution of $E_p$ in a GRB pulse (Deng \& Zhang 2014). Furthermore, the observed $E_{\rm p}$ of some GRBs (e.g. early emission of GRB 110721A) is beyond the ``death line'' of the photosphere model in the $L_{\rm iso}-E_{\rm p}$ plane (Zhang et al. 2012b), suggesting that a non-thermal origin of the Band component.

Interestingly, the correlation between jet power and prompt gamma-ray luminosity of GRBs is consistent with the correlation between jet power and synchrotron peak luminosity of FSRQs (Zhang et al. 2013a). This likely indicates that the radiation physics of the GRB Band function is analogous to that of the synchrotron peak in FSRQs. Notice that the most extensively studied GRB model is the synchrotron radiation from internal shocks of a baryonic fireball (e.g., M\'esz\'aros et al. 1994; Wang et al. 2009; Daigne et al. 2011). This model predicts the co-existence of a quasi-thermal photosphere component and a non-thermal synchrotron emission component in the observed spectra (M\'esz\'aros et al. 2000; Pe'er et al. 2006). Such a thermal emission component is either not detected or found sub-dominant in the broad band spectra for most GRBs observed with {\em Fermi} mission (e.g. Abdo et al. 2009; Guiriec et al. 2011; Axelsson et al. 2012), which suggests that the GRB jets are likely Poynting-flux-dominated (Zhang \& Pe'er 2009). Zhang \& Yan (2011) proposed that the pompt gamma-ray emission of GRBs may be originated through an internal-collision-induced magnetic reconnection and turbulence (ICMART) process. This model suggests that the majority of the Poynting flux energy is converted to the energies of electrons and protons efficiently, and the prompt emission is due to the synchrotron radiation of these electrons. The GRB emission radius of this model is generally large. Uhm \& Zhang (2014) showed that by considering decay of magnetic field strength in the emission region in an expanding jet, a fast-cooling synchrotron spectrum would mimic typical observed GRB Band functions. This process can also give rise to the observed GRB lightcurves (Zhang \& Zhang 2014).

The observed correlation between the peak photon energy and the corresponding luminosity may shed light on the radiation physics. Fossati et al. (1998) found a weak $L_{\rm s}-E_{\rm s}$ anti-correlation for blazars, which shows that high-luminosity FSRQs tend to have a low peak frequency and vice versa. This is the so-called ``blazar sequence'', which may be explained with more efficient cooling of particles in the jets of high-luminosity blazars (Ghisellini et al. 1998), or more physically, with the difference in BH masses and accretion rates among these sources (Ghisellini \& Tavecchio 2008). However, with a large blazar sample observed with {\em Fermi}/LAT, Meyer et al. (2011) found that the blazar sequence seems to break into two branches and form an envelope. In contrast to blazars, tight $E_{\rm iso}-E_{\rm p}$ (the Amati relation; Amati et al. 2002; Basak \& Rao 2013) and $L_{\rm iso}-E_{\rm p}$ (the Yonetoku relation; Yonetoku et al. 2004) correlations are observed among different GRBs and also within individual GRBs for different time slices (Liang et al. 2004; Lu et al. 2010, 2012), where $E_{\rm iso}$ and $L_{\rm iso}$ are the bolometric isotropic gamma-ray energy of a GRB and the bolometric isotropic luminosity at the peak time of a GRB lightcurve, respectively.

Recently, we have presented a systematical analysis of GRBs and blazars observed with {\em Fermi}/LAT in order to make a comparison of the relativistic jet properties in GRBs and blazars (Zhang et al. 2012a; 2013a,b). This paper presents a spectral analysis for a sample of redshift-known long GRBs observed with {\em Fermi}/GBM and makes a comparison of the distributions of GRBs and blazars (FSRQs, LBLs, IBLs, and HBLs) by plotting the synchrotron peak luminosity ($L_{\rm s}$) and the corresponding peak photon energy $E_{\rm s}$ of blazars in the $L_{\rm p}-E_{\rm p}$ plane of GRBs. We tentatively suggest a unified picture of radiation physics for the relativistic jets in GRBs and blazars within the framework of the leptonic ynchrotron radiation models. Our sample selection and data analysis processes are reported in \S 2, and the distributions of GRBs and blazars in the $L_{\rm  p}-E_{\rm p}$ plane are shown in \S 3 and \S 4, respectively. In \S 5, we discuss physical implications of the results within the framework of the leptonic synchrotron radiation models. Conclusions are presented in \S 6.  Throughout this work, we assume a flat $\Lambda$CDM universe with $\Omega_{\rm m}=0.3$, $\Omega_{\Lambda}=0.7$, $H_{0}=70\ {\rm km}\ {\rm s}^{-1}\ {\rm Mpc}^{-1}$.

\section{Data Analysis}

\subsection{GRB Data Reduction and Spectral Fits}
Redshift measurement is available for 33 long {\em Fermi} GRBs by March 2012. We download their data from the NASA {\em Fermi} web site\footnote{ftp://legacy.gsfc.nasa.gov/fermi/data/}. GBM has 12 sodium iodide (NaI) detectors covering an energy range from 8 keV to 1 MeV, and two bismuth germanate (BGO) scintillation detectors sensitive to higher energies between 200 keV and 40 MeV (Meegan et al. 2009). The signals from all the 14 GBM detectors are collected by a central Data Processing Unit, which packages the resulting data into three different types: CTIME, CSPEC, and TTE. The TTE event data files contain individual photons with time and energy tags. We use the TTE
data to make spectral fits with the software package RMFIT (version 3.3pr7). User-defined intervals before and after the prompt emission
phase are selected to obtain the background spectrum. We make a joint spectral fit to the spectra collected by the NaI and BGO
detectors with the Band function (Band et al. 1993). Our sample includes only those GRBs whose time-integrated spectra are well fitted with the Band function in the GBM band (8-1000 keV). GRBs 101219B, 111107,  and 110128A are excluded since their $E_{\rm p}$ values are poorly constrained with the data. We also exclude GRB 090902B since its spectrum is thermal emission dominated (Abdo et al. 2009b; Ryde et al. 2010; Zhang et al. 2011). We finally have a sample of 29 GRBs. Our spectral fitting results are reported in Table 1. Since the observed distribution of the high-energy photon index $\beta$ has a typical value of -2.25 (Preece et al. 2000), we fix $\beta$ to this value if the data quality is not good enough to constrain $\beta$.

\subsection {Blazar Samples}
We systematically model the observed SEDs for {\em Fermi}/LAT blazars to investigate their jet properties. We derive the physical parameters of the jets with the single zone leptonic model using the minimization $\chi^2$ technique for a sample of 24 TeV-selected BL Lacs and 23 Fermi/LAT bright FSRQs. Since two SEDs (low state and high state) are available for some BL Lacs in Zhang et al. (2012a, 2013), we actually obtain a sample of 57 well-sampled broadband SEDs for these sources. Our SED modeling results are reported in Zhang et al. (2012a, 2013a, b). We calculate $E_{\rm s}$ and $L_{\rm s}$ for these SEDs based on the model parameters reported in Zhang et al. (2012a, 2013a, b). The results are reported in Table 2.

Meyer et al. (2011) collected a large sample of blazars. We also select those {\em Fermi} blazars with confirmed redshift measurements in Meyer et al. (2011). We adopt the $E_{\rm s}$ and $L_{\rm s}$ of these blazars from Meyer et al. (2011), who utilized a phenomenological parametric model to fit the SEDs in order to estimate these parameters (Fossati et al. 1997). We also quote the data from Meyer et al. (2011) and list them in Table 2. Among them 145 are FSRQs and 71 are BL Lacs. Some TeV sources in Zhang et al. (2012, 2013) are also included in the blazar sample from Meyer et al. (2011).

\section{$L_{\rm p}-E_{\rm p}$ relation among GRBs}

We define $L_{\rm p}$ as the {\em mono-frequency}, isotropic luminosity at the spectral peak $E_{\rm p}$ of the {\em time integrated} spectrum of a GRB. This is different from most other luminosities adopted in previous studies. For example, the GRB luminosity adopted in the Yonetoku relation is the {\em bolometric} ($1-10^4$ keV) isotropic luminosity at the {\em peak time} of a GRB lightcurve, which we denote as $L_{\rm bol}^p$ in the following discussion.

\begin{figure}
\centering
\includegraphics[scale=0.6]{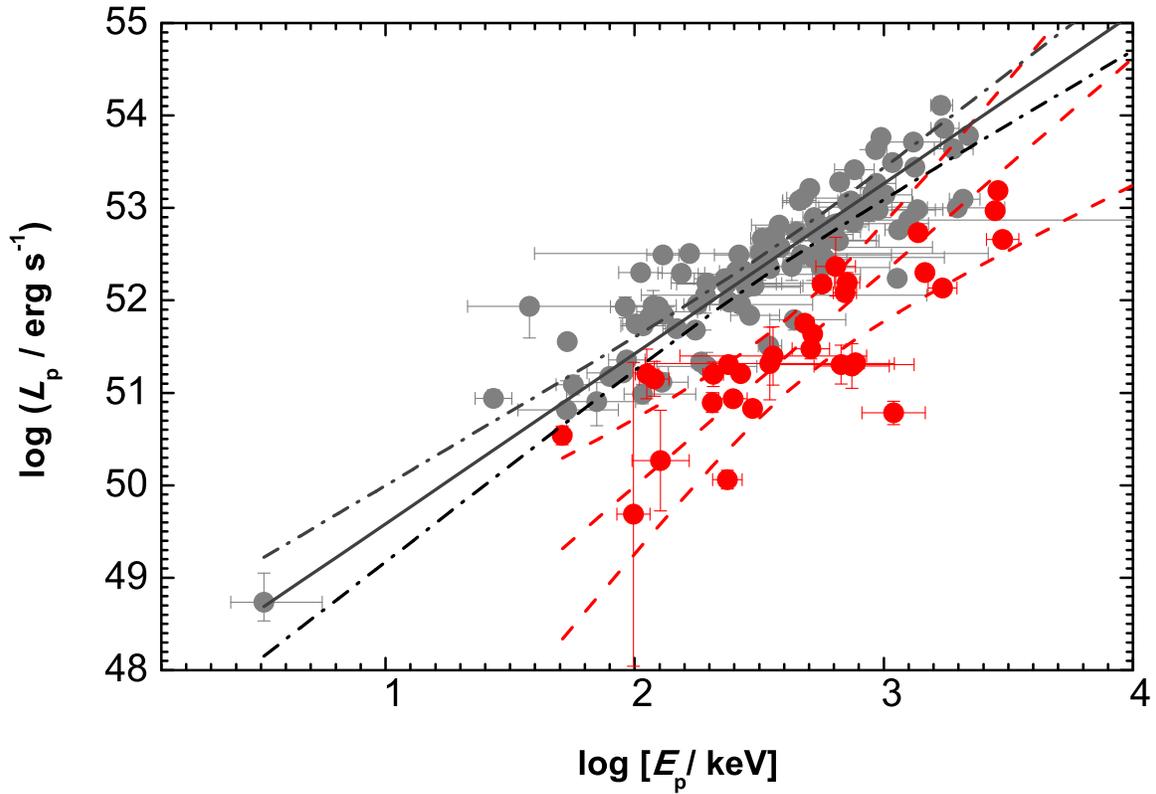}
\caption{Relation between $L_{\rm p}$ and $E_{\rm p}$ derived from the time-integrated spectra of GRBs in our sample. The Yonetoku relation (grey dots) is also shown with a sample of GRBs from Yonetoku et al. (2010). Lines are the linear fit and its 3 $\sigma$ confidence level to the data using the maximum likelihood method.} \label{Ep_Lp_GRB}
\end{figure}

In order to compare the GRBs and blazars in the $L_{\rm p}-E_{\rm p}$ plane, we first derive the mono-frequency $L_{\rm p}-E_{\rm p}$ relation for GRBs in the pure spectral domain, then show the distributions of blazars by plotting their $L_{\rm s}$ and $E_{\rm s}$ in the $L_{\rm p}-E_{\rm p}$ plane of GRBs. With the data reported in Table 1, we first study the $L_{\rm p}-E_{\rm p}$ relation for the GRBs in our sample. As shown in Fig.\ref{Ep_Lp_GRB}, the two quantities are tightly correlated. We make a linear fit to the data by considering the errors of the two quantities, and an intrinsic scatter by using the maximum likelihood method (e.g., D'Agostini 2005; Amati et al., 2008; Hogg et al. 2010). We get $\log L_{\rm p}=45.36^{+1.20}_{-1.47}+2.13^{+0.54}_{-0.46}\times \log E_{\rm _p}$ with an intrinsic scatter of $ 0.27^{0.06}_{-0.04}$.
We also show the Yonetoku relation with a GRB sample from Yonetoku et al. (2010) in Fig. \ref{Ep_Lp_GRB}. Our fit with the maximum likelihood method provides $\log L_{\rm bol}^p=47.75^{+0.42}_{-0.29}+1.84^{+0.11}_{-0.14}\times \log E_{\rm p}^p$ with an intrinsic scatter of $0.19^{+0.03}_{-0.01}$, where $E_{\rm p}^p$ is the peak energy at the peak time.
One can find that the slopes of the two relations are consistent with each other within the error bars. The $L_{\rm bol}^p$ value is systematically 1 order of magnitude larger than $L_{\rm p}$. This is reasonable in view of the definitions of the two luminosities.
Below we estimate the ratio $L_{\rm bol}^p/L_{\rm p}$ quantitatively. The Band function used for GRB spectral fits is
\begin{equation}
N(E)=N_{E}\times\cases{E^{\alpha}e^{-E/E_{\rm b}}, &
$E<(\alpha-\beta)E_{\rm b}$, \cr
[(\alpha-\beta)E_{\rm b}]^{\alpha-\beta}e^{\beta-\alpha}E^{\beta},&
$E\geq(\alpha-\beta)E_{\rm b}$,}
\end{equation}
with a typical low energy photon index $\alpha \sim -1.0$ and a high energy photon index $\beta \sim -2.2$. $E_{\rm p}$ is defined as the peak energy of $E^{2}N(E)$, and
\begin{equation}
E_{\rm p} = (2+\alpha)E_{\rm b},
\end{equation}
for $\alpha>-2.0$ and $\beta<-2.0$. The peak flux density is
therefore
\begin{equation}
F_{E_{\rm p}} = N_E[(2+\alpha)E_{\rm b}]^{1+\alpha}e^{-\alpha-2},
\end{equation}
while the total bolometric flux is $F = \int_{0}^{\infty} EN(E)dE$.
The luminosity ratio is thus $L_{\rm bol}^p/L_{\rm p} = F/(E_{\rm p}F_{E_{\rm p}})$, or
\begin{equation}
\displaystyle \frac{L_{\rm bol}^p}{L_{\rm p}} =
\displaystyle\left(\frac{\alpha-\beta}{2+\alpha}\right)^{2+\alpha}\left[e^{2+\alpha}\int_{0}^{1}x^{1+\alpha}e^{(\beta-\alpha)x}dx-\frac{e^{2+\beta}}{\beta+2}\right].
\end{equation}
For $\alpha\sim-1.5$ to $-0.5$ and $\beta\sim -2.2$ to $-2.1$, one has $L_{\rm bol}^p/L_{\rm p}\sim 6 - 13$, consistent with the above statistics. Therefore, $L_{\rm p}$ could be a good representative of $L_{\rm bol}^p$, and $E_{\rm p}$ of the time-integrated spectrum is dominated by the spectrum accumulated at the brightest time interval. These results indicate that the $L_p-E_p$ relation in the spectral domain could be the foundation of the Yonetoku relation.

\section{GRBs and blazars in the $L_{\rm p}-E_{\rm p}$ plane}
We add the blazars in our sample to the $L_{\rm p}-E_{\rm p}$ plane by plotting their $L_{\rm s}$ against $E_{\rm s}$ in Figure 2. It is found that GRBs, FSRQs together with LBLs, and IBL and HBLs occupy three different regions. GRBs are in the high-$L_{\rm p}$, high-$E_{\rm p}$ region, with a tight $L_{\rm p}-E_{\rm p}$ relation. Both IBLs and HBLs are in the range of $E_{\rm s}=2\times 10^{-3}\sim 10^{2}$ keV and $L_{\rm s}=10^{44}-10^{47}$ erg s$^{-1}$. No obvious dependence of $L_{\rm s}$ on $E_{\rm s}$ is found. The FSRQs and LBLs are clustered  in the region of  $E_{\rm s}=10^{-5}\sim 2\times 10^{-3}$ keV and $L_{\rm s}=10^{44}\sim 10^{48}$ erg s$^{-1}$. A weak, anti-correlation trend is seen in FSRQs or LBLs, with a lower $E_{\rm s}$ corresponding to a higher $L_{\rm s}$. However, our correlation analysis cannot conclude any statistically significant correlation between the two quantities for these sources.
\section{Implications of the radiation physics}
As is shown above, the distributions of  GRBs, FSRQs together with LBLs, and IBL and HBLs in the $L_{\rm p}-E_{\rm p}$ plane are different and the relation between the two quantities are also dramatically different. In the following, we discuss the possible reasons that may shape the GRB and blazar distributions in the $L_{\rm p}-E_{\rm p}$ plot in the framework of the synchrotron radiation mechanism. In the cosmic proper rest frame, the synchrotron radiation peaks at
\begin{equation}
E_{\rm p} \sim \hbar \delta \gamma_{\rm e,p}^2 \frac{e B'}{m_e c}~,
\label{Epsyn}
\end{equation}
where $\hbar$ is the Plank constant, $\delta$ is the Doppler factor of the radiating region, $\gamma_{\rm e,p}$ is the Lorentz factor of the electrons responsible for the radiation at $E_{\rm p}$, $m_e$ and $e$ are the electron mass and charge, $c$ is the speed of light, $z$ is the redshift, and $B'$ is the magnetic field in the comoving frame. The value of $E_{\rm p}$ strongly depends on $\gamma_{\rm e, p}$, $\delta$, and $B^{'}$. Various observations suggest that the GRB Doppler factor ranges from several 10s to above 1000 (Liang et al. 2010; Racusin et al. 2011; L\"u et al. 2012 and references therein). The $L_{\rm s}$ values of blazars are significantly lower than GRBs and are distributed in the range of $\sim 10^{44} - 10^{48}$ erg s$^{-1}$. Their Doppler boosting factors are also lower. As reported by Zhang et al. (2013), the distribution of the $\delta$ values of both BL Lacs and FSRQs are similar, which range from a few to a few tens. The $E_{\rm s}$ of BL Lacs has a broad distribution in the range of $10^{-5}\sim 10^2$ keV, but the $E_{\rm s}$ of FSRQs are usually lower than $10^{-3}$ keV. The broad $E_{\rm s}$ distribution of BL Lacs may result from the broad distribution of $\gamma_{\rm e,p}$ among different sources. In the following, we discuss the dependence of $E_{\rm s}$ on various physical parameters and jet composition for different jets.

\subsection{Synchrotron Radiation of Electrons in the Fast Cooling Regime for GRBs}
The shape of synchrotron radiation spectrum depends on the relative order of two characteristic frequencies, the minimum frequency $\nu_m$ (corresponding to emission from the minimum injection Lorentz factor $\gamma_m$ of electrons) and the cooling frequency $\nu_c$. The regimes of fast cooling and slow cooling correspond to $\nu_c < \nu_m$ and $\nu_m < \nu_c$, respectively (Sari et al. 1998). Due to their extremely high luminosities, the magnetic field strength in the emission region of GRBs is very high, so that fast cooling applies.

The most extensively discussed GRB model is the baryonic fireball model. The internal energy of the fireball is assumed to be distributed among protons, electrons and magnetic fields with the energy partition fractions $\epsilon_{ p}$, $\epsilon_{ e}$ and $\epsilon_{ B}$, respectively. In the case of that the radiation is from the synchrotron radiation of electrons in the fast cooling regime, one has (adapted from Eq.(8) of Zhang \& Yan 2011),
\begin{eqnarray}
E_{\rm s}& \simeq & \frac{2.78\times 10^{-2}}{1+z}~{\rm keV}~ \psi^2(p)
L_{\rm s,52}^{1/2} R_{14}^{-1}\epsilon_{B,-2}^{1/2}\epsilon_{e,-1}^{3/2}\epsilon_p^{-2} {(n_p/n_e)}^{2}(\bar\gamma_p-1)^2~,
\label{Epsyn-IS}
\end{eqnarray}
where $\psi(p)\equiv 6(p-2)/(p-1)$ is a function of the power law index of electron energy distribution $p>2$, which is $\sim 1$ for $p=2.2$; $L_{\rm s}$ is the synchrotron peak luminosity; $R$ is the radius of the emission region; $n_{e}$ and $n_{p}$ are the number densities of the electrons and protons, respectively; and $\bar\gamma_p$ is the mean random Lorentz factor of shock-accelerated protons in the comoving frame of the internal shocks, typically with $(\bar\gamma_p-1) \sim 1$ (Zhang \& Yan 2011). The notation $Q_{n}$ is defined as $Q/10^{n}$ in the cgs units. One can find that $E_{\rm s}$ depends on various parameters. It is difficult to predict a tight $L_{\rm s}-E_{\rm s}$ relation as observed in GRBs since these parameter values are dramatically different among GRBs. Therefore, the observed tight  $L_{\rm p}-E_{\rm p}$ relation may disfavor this scenario for GRBs.

The non-detection of photospheric emission in most GRBs (e.g., Zhang et al. 2011) and the detection of relatively high linear polarization in the prompt emission phase and the early reverse shock afterglow phase of several GRBs (Mundell et al. 2007, 2013, Steele et al.2009) suggest that the GRB ejecta may be magnetically dominated. Assuming that the ejecta is strongly magnetized and the internal energy through magnetic dissipation is distributed to electrons and protons in the fractions of $\varepsilon_{\rm e}$ and $\varepsilon_{\rm p}$ with $\varepsilon_e+\varepsilon_p=1$. The average Lorentz factor of the accelerated electrons in the ejecta is given by (Eq. (55) of Zhang \& Yan 2011)
\begin{equation}\label{gamma_e}
\bar\gamma_e=\eta\varepsilon_e (1+\sigma)(1+m_p/Y_e m_e)~,
\end{equation}
where $\sigma$ is the magnetization parameter, $\eta$ is the efficiency of the magnetic energy converted into internal energy, and $Y_e$ is the lepton (pair) multiplity
parameter. In the following, we assume $Y_e\ll m_p/m_e$.
The minimum injection Lorentz factor is $\gamma_m = [(p-2)/(p-1)] \bar \gamma_e$.

Since the total ``wind" luminosity is
\begin{equation}\label{Lwind}
L_{\rm w}=\frac{\sigma+1}{\sigma}4\pi R^{2}\frac{B'^{2}}{8\pi}\Gamma^{2}c,
\end{equation}
and the co-moving magnetic field can be estimated as
\begin{equation}\label{B}
B'=\left[\frac{2L_{w}\sigma}{(1+\sigma) R^{2}\Gamma^{2}c}\right]^{\frac{1}{2}},
\end{equation}
where $\Gamma$ is the bulk Lorentz factor of the radiating jet. We assume that for GRBs the moving direction of the relativistic jet is along the line of sight, so the Doppler boosting factor $\delta\sim \Gamma$. In the case of strong magnetization, the electron cooling is dominated by the synchrotron radiation rather than inverse Compton scattering.

The fast cooling regime corresponds to $\gamma_c < \gamma_m$.
In this case, the radiation at the synchrotron peak is attributed to the electrons with minimum injection Lorentz factor. The synchrotron radiation luminosity in the fast cooling case can be estimated as $L_{\rm s}\simeq \varepsilon_e \eta L_{w}$. This case was discussed in detail in Zhang \& Yan (2011). The expected $L_{\rm s}-E_{\rm s}$ relation can be quoted as following for GRBs,
\begin{eqnarray}\label{Es_fast}
E_{\rm s}& \simeq 320~{\rm keV}~ \psi^2(p)
\eta^{3/2} \varepsilon^{3/2}_{e} Y^{-2} L_{\rm s,52}^{1/2} R_{_{\rm 15}}^{-1}
 \sigma_{1.5}^{2}(1+z)^{-1}.
\end{eqnarray}

This scenario may predict a typical $E_{\rm p}$ of GRBs, and one can also find an implicit dependence of $E_{\rm s}\propto L_{\rm s}^{1/2}$. However, a tight $E_{\rm s}\propto L_{\rm s}^{1/2}$ relation requires that other parameters are almost universal among GRBs or they are independent on $L_s$. The value of $\varepsilon_e$ depends on the particle acceleration mechanism. It is generally believed that electrons in GRB jets are accelerated via the Fermi acceleration mechanism. For a given acceleration mechanism, $\varepsilon_e$ may not vary significantly. For the ICMART model (Zhang \& Yan 2011), a higher $\sigma$ may result in a larger $R$ for energy dissipation to happen, which would cancel out the $R-$ and $\sigma-$ dependences. As a result, a Yonetoku-like relation is expected for GRBs within the fast cooling synchrotron radiation scenario in a strongly magnetized outflow.

No clear $E_{\rm s}-L_{\rm s}$ dependency is found for the  entire blazar sample. The observed $E_{\rm s}$ of blazars range from $\sim 10^{-5}$ to 100 keV and the typical value of $L_{\rm s}$ is $\sim 10^{44} - 10^{48}$ erg s$^{-1}$. The typical $R$ estimated with the observed variability timescale is $\sim 10^{15}$ cm. For FSRQs, the jet radiation efficiency $\varepsilon = \varepsilon_e\eta \sim 0.1$ and $\sigma\sim 0.1 - 10$ (Zhang et al. 2013a), one can estimate the synchrotron peak energy $E_{\rm s} \simeq 10^{-5}~{\rm keV}~ \psi^2(p) \varepsilon^{3/2}_{-1} Y^{-2} L_{\rm s,46}^{1/2} R_{_{\rm 15}}^{-1}
 \sigma^{2}(1+z)^{-1}$. The estimated typical value of $E_{\rm s}$ is smaller than the observed one for FSRQs, and the trend of $E_{\rm s}$ with $L_{\rm s}$ is also inconsistent with the observations for FSRQs. For BL Lacs, the jet radiation efficiency $\varepsilon \sim 10^{-4} - 0.1$ and $\sigma\sim 10^{-4} - 1$ (Zhang et al. 2013a), one can estimate the synchrotron peak energy $E_{\rm s} \simeq 10^{-7}~{\rm keV}~ \psi^2(p) \varepsilon^{3/2}_{-2} Y^{-2} L_{\rm s,46}^{1/2} R_{_{\rm 15}}^{-1}
 \sigma^{1/2}_{-1}(1+z)^{-1}$. The estimated typical value of $E_{\rm s}$ is much smaller than the observed one for BL Lacs. In conclusion, the observations should disfavor the fast cooling scenario for blazars. This is consistent with the theoretical expectation: the magnetic field strength in the blazar emission region is low, so that usually slow cooling should apply.

\subsection{Synchrotron Radiation of Electrons in the Slow Cooling Regime for blazars}

The slow cooling regime corresponds to $\gamma_c > \gamma_m$. In this case, the radiation at the synchrotron peak should be attributed to the electrons with a Lorentz factor $\gamma_e\sim \gamma_c$, which is given by \begin{equation}
\gamma_{c}=\frac{6\pi m_{e}c}{\sigma_{T}B'^{2}t'},
\end{equation}
where $t^{'}$ is the cooling time scale in the comoving frame and $\sigma_{\rm T}$ is the Thomson cross section. We assume that $t^{'}$ is comparable to the dynamic timescale, i.e., $R=\Gamma c t'$. The synchrotron luminosity in the slow cooling case can be estimated as $L_{\rm s}\simeq \varepsilon_{\rm rad}\varepsilon_e \eta L_{w}$, where $\varepsilon_{\rm rad}$ is the fraction of the electron energy that is radiated away. The radiation efficiency of electrons in the slow cooling can be estimated as $\varepsilon_{\rm rad} \simeq (\bar{\gamma}_e/\gamma_c)^{p-2}$ (Sari \& Esin 2001). The jet radiation efficiency $\varepsilon = \varepsilon_{\rm rad}\varepsilon_e\eta \sim 0.1$ is estimated for FSRQs (Zhang et al. 2013a). For a general discussion, we derive the cooling Lorentz factor of electrons for an outflow with an arbitrary $\sigma$:
\begin{eqnarray}{\label{gamma_c}}
\gamma_{c} & = &\frac{3\pi m_{e}c^{3}(1+\sigma) R\Gamma^{3} \varepsilon}{\sigma_{T}\sigma L_{\rm s}} \nonumber \\
           & \simeq & 10^2 \varepsilon_{-1} (1+\sigma)\sigma^{-1}R_{15} \Gamma_{1.5}^{3} L_{\rm s, 46}^{-1}.
\end{eqnarray}
Therefore, $E_{\rm s}$ can be written as
\begin{eqnarray}\label{Es_slow}
E_{\rm s}&\simeq&\delta\gamma_{c}^{2}\hbar \frac{eB'}{
m_{e}c}(1+z)^{-1}\nonumber \\
&=& 3.2\times 10^{-4}{\rm keV}L^{-3/2}_{\rm s,46}\Gamma_{1.5}^{6} R_{15}\varepsilon_{-1}^{3/2}(\frac{1+\sigma}{\sigma})^{3/2}(1+z)^{-1}.
\end{eqnarray}

\begin{figure*}
\centering
\includegraphics[angle=0,scale=0.6]{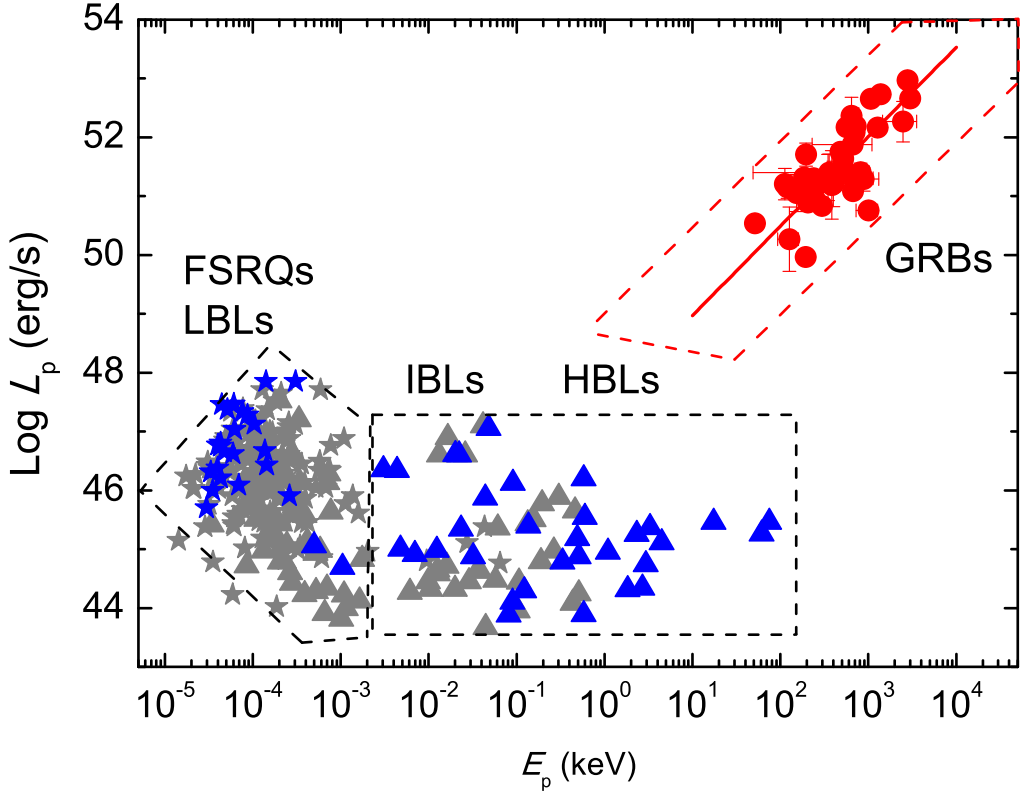}
\caption{Distributions of the GRBs and blazars in the $L_{\rm p}$ and $E_{\rm p}$ plane, in which the red dots are for the GRBs, the stars are for the FSRQs, and triangles are for the BL Lacs. The line is the best fit to the GRB data. The grey data points of blazars are taken from Meyer et al. (2011) and the other data points of blazars are taken from Zhang et al. (2012a, 2013). } \label{eplp_new}
\end{figure*}

For a moderate $\sigma \sim 1$ which is relevant for FSRQs (Zhang et al. 2013a), one can see that the typical $E_{\rm s}$ is consistent with that observed in FSRQs and LBLs. An apparent $E_{\rm s}-L_{\rm s}$ anti-correlation is expected from Eq. (\ref{Es_slow}), which seems to be consistent with the global distirbution of FSRQs and LBLs in the $L_{\rm p}-E_{\rm p}$ domain, as shown in Figure \ref{eplp_new}. However, $E_{\rm s}$ strongly depends on $\Gamma$ and other parameters like $\sigma$, $\varepsilon$). These parameters vary among different sources (e.g. Zhang et al. 2012a, 2013b). One therefore would not expect a tight $E_{\rm s}-L_{\rm s}$ anti-correlation. Within a same blazar, given a black hole with the same mass and spin, one would expect that the variation mostly depends on the Doppler factor. Theoretical arguments and observational data suggest that generally one has $L_s\propto \delta^{3-4}$ (e.g. Ghisellini et al. 1998; Wu et al. 2011; Zhang et al. 2013a). Submitting this dependence to Eq.(\ref{Es_slow}), one gets $E_{\rm s}\propto L^{0-0.48} R\varepsilon^{3/2}(\frac{1+\sigma}{\sigma})^{3/2}$.  It is interesting that Zhang et al. (2012) showed a tentative correlation $L_s\propto E_s^{0.28- 0.45}$ within individual blazars in different flux states. This is fully consistent with the theoretical framework proposed in this paper.

For IBLs and HBLs, the jet composition is likely matter dominated (Ghisellini et al. 2010; Zhang et al. 2013a). With $\sigma << 1$, $E_s$ (Eq.(\ref{Es_slow})) is expected to be very sensitive to $\sigma$, and the rough anti-correlation between $E_s$ and $L_s$ is expected to be destroyed by this sensitive dependence. This naturally explains the spread of IBL and HBL along the $E_s$ axis in the $L_{\rm p} - E_{\rm p}$ plane.

\section{Conclusions}
We have derived a tight $L_{\rm p}-E_{\rm p}$ relation from the time-integrated spectra of {\em Fermi} GRBs, i.e., $L_{\rm p}\propto E_{\rm p}^{2.13^{+0.54}_{-0.46}}$. We show that this correlation is physically related to the Yonetoku relation and the Amati relation. We compare the {\em Fermi} blazars with GRBs by plotting their $L_{\rm s}$ and $E_{\rm s}$ in the $L_{\rm p}-E_{\rm p}$ plane of GRBs. The GRBs are in the high-$L_{\rm p}$, high-$E_{\rm p}$ corner of the plane, illustrating as a tight $L_{\rm p}-E_{\rm p}$ relation. Both FSRQs and LBLs are clustered in the low-$E_{\rm p}$, low-$L_{\rm p}$ corner with a weak anti-correlation trend. IBLs and HBLs range in $E_{\rm s}\sim 2\times 10^{-3} - 10^{2}$ keV and $L_{\rm s}=10^{44}\sim 10^{47}$ erg s$^{-1}$, but no dependence of $L_{\rm s}$ on $E_{\rm s}$ is found.

We discuss the possible reasons that may shape the distributions of GRBs and blazars in the $L_{\rm p}-E_{\rm p}$ plane within the framework of leptonic synchrotron radiation models. It seems that a self-consistent, unified picture is available to account for all the data. For GRBs, thanks to their extremely high luminosity, the co-moving frame magnetic field strength is so high that electrons are in the fast cooling regime. This naturally gives rise to a positive dependence between $L_{\rm p}$ and $E_{\rm p}$. In order to achieve a tight correlation, a magnetically dominated outflow is favored. For blazars, since their luminosities are much lower, electrons are likely in the slow cooling regime. This gives rise to a rough anti-correlation between $L_s$ and $E_s$, with sensitive dependence on $\Gamma$, and also on $\sigma$ when $\sigma << 1$. FSRQs and LBLs can maintain a rough anti-correlation, suggesting that they have a moderately high $\sigma$. This is consistent with the modeling of Zhang et al. (2013a). IBLs and HBLs are matter dominated with $\sigma << 1$. The sensitive dependence of $E_s$ on $\sigma$ therefore introduces a wide spread of these objects in the $L_{\rm p}-E_{\rm p}$ plane. Our results indicate that there is a unified mechanism for relativistic jets. Different observational appearances can be attributed to different radiation regimes (fast vs. slow cooling) and different jet compositions.

This work is supported by the National Basic Research Program (973 Programme) of China (Grant 2014CB845800), the National Natural Science Foundation of China (Grants 11025313, 11363002, 11322328), Guangxi Science Foundation (2013GXNSFFA019001), the Strategic Priority Research Program ¡°The Emergence of Cosmological Structures¡± of the Chinese Academy of Sciences (Grant No. XDB09000000), and the Key Laboratory for the Structure and Evolution of Celestial Objects of Chinese Academy of Sciences. X. F. Wu acknowledges support by the One-Hundred-Talents Program, the Youth Innovation
Promotion Association and the
Natural Science Foundation of Jiangsu Province.

\begin{deluxetable}{lccccccc}
\centering
\tabletypesize{\scriptsize}
\tablewidth{440pt}
 \tablecaption{Results of Our Spectral Fits with the Band function to the Time-Integrated spectra for the GRBs in our sample\label{tb-spp}}
 \tablehead{
 \colhead{GRB} & \colhead{$z$} & \colhead{$A$} &
\colhead{$\alpha$}  & \colhead{$\beta$} &\colhead{$E_{\rm p}$}&
\colhead{$\chi^{2}$/dof} & \colhead{ $L_{\rm p}$}
\\

\colhead{} & \colhead{} & \colhead{phs/cm$^{2}{\cdot}s{\cdot}$keV} &
\colhead{}  & \colhead{} & \colhead{(keV)}  & \colhead{}&
\colhead{(erg/s)}
}
 \startdata
080905B&2.374&0.00276$\pm$0.00080&-1.186$\pm$0.193&-2.250\tablenotemark{a}&263.0 $\pm$128.0 &1.089&51.289$\pm$0.203 \\
080916C&4.35&0.01583$\pm$0.00038&-1.037$\pm$0.017&-2.251$\pm$0.135&523.3 $\pm$36.4 &1.210&52.968$\pm$0.028 \\
080916A&0.689&0.00643$\pm$0.00128&-1.032$\pm$0.106&-2.250&114.6 $\pm$15.0 &0.999&49.969$\pm$0.111 \\
081121&2.512&0.02266$\pm$0.00340&-0.586$\pm$0.109&-2.044$\pm$0.088&199.0 $\pm$21.5 &1.146&52.083$\pm$0.088 \\
081222&2.77&0.02840$\pm$0.00289&-0.833$\pm$0.058&-2.211$\pm$0.107&149.9 $\pm$11.9 &1.130&52.175$\pm$0.065\\
090323&3.57&0.00999$\pm$0.00048&-1.034$\pm$0.041&-2.418$\pm$0.314&656.5 $\pm$98.8 &1.178&52.659$\pm$0.068 \\
090423&8.1&0.00670$\pm$0.00386&-1.001$\pm$0.296&-2.250$\pm$0.000&70.4 $\pm$12.8 &0.971&52.366$\pm$0.318 \\
090424&0.544&0.14150$\pm$0.00368&-0.917$\pm$0.016&-2.651$\pm$0.078&172.5 $\pm$3.7 &1.956&51.210$\pm$0.016 \\
090516A&4.109&0.00411$\pm$0.00029&-1.085$\pm$0.049&-2.250&337.1 $\pm$44.8 &1.181&52.135$\pm$0.061 \\
090618&0.54&0.04678$\pm$0.00123&-1.311$\pm$0.017&-2.566$\pm$0.087&193.3 $\pm$6.5 &1.469&50.833$\pm$0.014\\
090926B&1.24&0.04058$\pm$0.00800&-0.155$\pm$0.114&-3.438$\pm$0.589&91.4 $\pm$4.1 &1.060&50.895$\pm$0.107 \\
090927&1.37&0.00914$\pm$0.01490&-0.636$\pm$0.805&-2.250&53.6 $\pm$14.2 &0.982&50.267$\pm$0.546 \\
091003&0.8969&0.02244$\pm$0.00081&-1.049$\pm$0.027&-2.252$\pm$0.152&405.9 $\pm$33.6 &1.137&51.325$\pm$0.035 \\
091020&1.71&0.00649$\pm$0.00165&-1.364$\pm$0.145&-1.844$\pm$0.140&249.0 $\pm$168.0 &1.031&51.305$\pm$0.210 \\
091024&1.091&0.00320$\pm$0.00034&-1.058$\pm$0.078&-2.250&480.0 $\pm$132.0 &0.963&50.759$\pm$0.110 \\
091127&0.49&0.08341$\pm$0.01620&-1.393$\pm$0.078&-2.203$\pm$0.021&34.4 $\pm$2.0 &1.385&50.539$\pm$0.103 \\
091208B&1.063&0.01559$\pm$0.00199&-1.297$\pm$0.067&-2.507$\pm$0.391&120.3 $\pm$15.7 &1.093&50.936$\pm$0.067\\
100414A&1.368&0.03356$\pm$0.00081&-0.373$\pm$0.035&-2.250&556.7 $\pm$20.6 &1.278&52.968$\pm$0.110\\
100704A&3.6&0.01739$\pm$0.00323&-0.575$\pm$0.114&-2.304$\pm$0.234&155.2 $\pm$18.1 &1.055&52.192$\pm$0.109 \\
100728B&2.106&0.00866$\pm$0.00153&-1.115$\pm$0.104&-2.250&164.2 $\pm$28.4 &1.139&51.475$\pm$0.107 \\
100814A&1.44&0.03433$\pm$0.00862&-0.704$\pm$0.127&-2.404$\pm$0.170&84.7 $\pm$8.0 &1.002&51.201$\pm$0.133 \\
100816A&0.8035&0.14640$\pm$0.02300&-0.256$\pm$0.095&-2.442$\pm$0.150&131.8 $\pm$8.3 &1.151&51.307$\pm$0.087 \\
100906A&1.727&0.02642$\pm$0.00198&-1.001$\pm$0.043&-2.126$\pm$0.094&176.5 $\pm$14.5 &1.237&51.753$\pm$0.049 \\
110213A&1.46&0.02946$\pm$0.01070&-1.123$\pm$0.153&-2.137$\pm$0.066&48.8 $\pm$6.6 &1.123&51.152$\pm$0.189\\
110731A&2.83&0.03617$\pm$0.00157&-0.932$\pm$0.034&-2.302$\pm$0.137&359.0 $\pm$26.5 &1.172&52.733$\pm$0.041 \\
110818A&3.36&0.00486$\pm$0.00598&-0.955$\pm$0.536&-1.659$\pm$0.070&82.5 $\pm$71.3 &0.980&51.399$\pm$0.316 \\
120119A&1.728&0.01879$\pm$0.00086&-0.990$\pm$0.028&-2.534$\pm$0.171&190.0 $\pm$9.1 &1.409&51.636$\pm$0.027 \\
120326A&1.798&0.06001$\pm$0.03670&-0.635$\pm$0.252&-2.331$\pm$0.102&40.0 $\pm$4.3 &1.064&51.205$\pm$0.267 \\
\enddata
\tablenotetext{a}{$\beta$ is fixed at -2.25 when data cannot give well constrained while fitting.}

\end{deluxetable}

\begin{deluxetable}{lccc}
\centering
\tablewidth{400pt}
\tablecaption{Data of our blazar samples taken from Zhang et al. (2012a, 2013) and Meyer et al. (2011).\label{tb-spp}}
\tablehead{
\colhead{Name\tablenotemark{a}} & \colhead{$z$} & \colhead{log$[E_{\rm s}(1+z) ({\rm keV})]$} & \colhead{log [$L_{\rm s}$(erg/s)]}}
\startdata
TeV BL Lacs&&&\\
\hline
Mkn 421$^{\rm L}$&0.031&-0.295 &44.87 \\
Mkn 421$^{\rm H}$&&0.519 &45.37 \\
Mkn 501$^{\rm L}$&0.034&-0.914 &44.29 \\
Mkn 501$^{\rm H}$&&1.871 &45.46 \\
W Com$^{\rm L}$&0.102&-2.159 &44.91 \\
W Com$^{\rm H}$&&-2.320 &45.00 \\
BL Lacertae $^{\rm L}$&0.069&-3.304 &45.06 \\
BL Lacerta$^{\rm H}$&&-2.979 &44.69 \\
PKS 2005-489$^{\rm L}$ &0.071&-1.910 &44.98 \\
PKS 2005-489$^{\rm H}$&&-1.633 &45.34 \\
1ES 1959+650$^{\rm L}$&0.048&0.038 &44.95 \\
1ES 1959+650$^{\rm H}$&&1.786 &45.26 \\
1ES 2344+514$^{\rm L}$&0.044&-1.045 &44.10 \\
1ES 2344+514$^{\rm H}$&&0.432 &44.34 \\
PKS 2155-304 $^{\rm L}$&0.116&-1.354 &45.87 \\
PKS 2155-304 $^{\rm H}$&&-1.041 &46.12 \\
1ES 1101-232$^{\rm L}$&0.186&-0.223 &45.54 \\
1ES 1101-232$^{\rm H}$&&-0.308 &45.18 \\
S5 0716+714$^{\rm L}$&0.26&-2.517 &46.35 \\
S5 0716+714$^{\rm H}$&&-2.359 &46.34 \\
3C66A&0.44&-1.653 &46.60 \\
PG1553+113&0.3&-1.696 &46.61 \\
1ES 1218+30.4&0.182&-0.867 &45.39 \\
1ES 1011+496&0.212&-0.236 &46.20 \\
PKS 1424+240&0.5&-1.319 &47.05 \\
RGB J0710+591&0.125&0.651 &45.11 \\
1ES 0806+524&0.138&-1.496 &44.87 \\
Mkn 180&0.045&-1.092 &43.88 \\
H2356-309&0.165&-0.479 &44.78 \\
1ES 0347-121&0.188&0.365 &45.25 \\
1ES 0229+200&0.14&1.241 &45.46 \\
RGB J0152+017&0.08&-0.236 &43.89 \\
H1426+428&0.129&0.472 &44.74 \\
PKS 0548-322&0.069&0.263 &44.31 \\
\hline
TeV FSRQs&&&\\
\hline
3C279&0.536&-4.48&46.31\\
3C273&0.158&-4.16&46.09\\
3C454.3&0.859&-4.35&47.46\\
PKS 1454-354&1.424&-4.06&47.28\\
PKS 0208-512&1.003&-4.37&46.75\\
PKS 0254-234&1.003&-4.37&46.80\\
PKS 0727-11&1.589&-4.21&47.03\\
PKS 0528+134&2.07&-4.29&47.37\\
4C66.20&0.657&-4.40&46.37\\
4C29.45&0.729&-3.84&46.43\\
PKS0420+31&1.487&-4.23&46.62\\
PKS0420-01&0.916&-3.86&46.67\\
1Jy 1308+326&0.997&-4.37&46.21\\
PKS1510-089&0.36&-4.52&45.70\\
4C28.07&1.213&-4.32&46.65\\
PMN 2355-1555&0.621&-4.46&46.00\\
S3 2141+17&0.213&-3.58&45.91\\
S4 0133+47&0.859&-4.40&46.77\\
S4 0917+44&2.19&-4.12&47.34\\
PKS 0347-369&2.115&-3.98&47.12\\
PKS 0347-211&2.944&-3.85&47.85\\
PKS2325+093&1.843&-3.51&47.85\\
PKS 1502+106&1.839&-4.21&47.47\\
\hline
BL Lacs from Meyer et al &&&\\
\hline
QSO B1028+511&0.36&-0.52&45.88\\
QSO B1011+496&0.2&-2.75&44.83\\
QSO B0954+65&0.368&-3.52&45.82\\
QSO B0912+297&0.101&-1.98&44.32\\
QSO B0851+202&0.306&-3.77&46.09\\
QSO B0829+047&0.174&-3.70&45.43\\
QSO B0818-128&0.074&-3.18&43.91\\
QSO B0814+42&0.245&-4.07&44.72\\
QSO B0808+019&1.148&-3.66&46.36\\
QSO B0806+524&0.138&-2.01&44.45\\
QSO B0754+10&0.266&-3.53&45.67\\
QSO B0548-322&0.069&-0.29&44.24\\
QSO B0537-441&0.896&-3.86&46.94\\
QSO B0521-365&0.055&-3.28&44.28\\
QSO B0426-380&1.11&-4.06&46.55\\
QSO B0422+004&0.31&-3.51&45.91\\
QSO B0235+16&0.524&-3.96&46.26\\
QSO B0212+73&2.367&-3.48&47.2\\
QSO B0138-097&0.733&-3.65&46.22\\
QSO B0118-272&0.557&-3.15&46.13\\
QSO B0109+224&0.265&-3.41&45.57\\
QSO B0048-09&0.2&-3.87&44.97\\
7C 1055+5644&0.144&-1.80&44.71\\
CSO 916&0.108&-1.24&44.47\\
BWE 1705+7142&0.208&-1.54&44.45\\
BWE 1413+4844&0.496&-3.38&45.13\\
7C 0937+2617&0.498&-3.76&45.27\\
BWE 0812+5748&0.294&-1.44&44.63\\
B3 1747+433&0.215&-3.41&44.23\\
8C 2007+777&0.342&-3.96&45.51\\
8C 1803+784&0.684&-3.52&46.53\\
8C 1148+592&0.118&-2.22&44.27\\
8C 0716+714&0.3&-3.14&46.28\\
7C 1724+5854&0.125&-2.94&43.99\\
7C 1415+2556&0.237&-0.72&44.79\\
7C 1308+3236&0.997&-4.02&46.69\\
7C 1219+2830&0.1&-3.61&45.07\\
7C 1147+2434&0.2&-3.28&44.93\\
4C 47.08&0.475&-3.46&46.13\\
4C 39.49&0.034&-1.42&44.63\\
4C 22.21&0.951&-4.00&45.7\\
4C 15.54&0.357&-3.69&44.78\\
4C 09.57&0.322&-4.08&45.74\\
3C 446&1.404&-3.69&47.52\\
3C 371&0.051&-2.96&44.22\\
VIPS 1105&0.146&-2.78&44.12\\
\hline
FSRQ from Meyer et al &&&\\
\hline
4C 72.16&1.46&-3.99&46.25\\
4C 71.07&2.172&-3.97&47.18\\
4C 58.17&1.318&-3.17&46.05\\
4C 56.27&0.663&-4.11&46.15\\
4C 55.17&0.895&-3.55&46.02\\
4C 53.28&0.98&-3.51&45.62\\
4C 51.37&1.375&-3.61&46.85\\
4C 49.22&0.334&-4.22&45.43\\
4C 47.44&0.74&-3.89&45.19\\
4C 47.29&1.462&-3.87&46.75\\
4C 47.23&1.292&-2.96&46.88\\
4C 40.25&1.254&-2.80&45.61\\
4C 40.24&1.252&-4.41&46.34\\
4C 39.25&0.695&-4.51&46.42\\
4C 38.41&1.814&-3.51&47.12\\
4C 32.14&1.258&-3.56&46.57\\
4C 31.63&0.298&-4.03&45.7\\
4C 29.45&0.729&-3.65&46.29\\
4C 28.07&1.207&-4.05&46.96\\
4C 23.24&0.565&-3.71&45.85\\
4C 21.35&0.435&-3.45&45.45\\
4C 20.24&1.11&-4.08&46.01\\
4C 19.34&0.828&-3.63&45.47\\
4C 15.05&0.405&-4.85&45.15\\
4C 14.60&0.605&-3.58&45.77\\
4C 13.14&2.059&-3.81&47\\
4C 12.39&2.129&-3.80&46.36\\
4C 11.69&1.037&-3.79&46.45\\
4C 10.45&1.226&-4.20&46.81\\
4C 06.69&0.99&-4.26&46.91\\
4C 06.41&1.27&-4.23&47.04\\
4C 06.11&0.511&-3.80&45.64\\
4C 05.64&1.422&-3.96&46.55\\
4C 04.42&0.965&-4.39&46.03\\
4C 01.28&0.888&-4.06&46.61\\
4C 01.24&1.024&-3.24&46.49\\
4C 01.02&2.099&-3.89&47.71\\
4C -03.79&0.901&-4.05&46.52\\
4C -02.81&1.285&-3.93&46.76\\
4C -02.19&2.286&-3.78&46.98\\
4C +45.30&0.495&-3.81&45.34\\
3C 66A&0.444&-3.24&46.34\\
3C 454.3&0.859&-3.58&46.88\\
3C 345&0.593&-3.93&46.5\\
3C 279&0.536&-4.33&46.71\\
3C 273&0.158&-3.40&45.86\\
3C 245&1.029&-4.45&46.24\\
3C 216&0.67&-3.96&45.8\\
\enddata
\tablenotetext{a}{Sources marked with ``H'' or ``L" indicate the high and low states as defined in Zhang et al. (2012a).}

\end{deluxetable}


\end{document}